\documentclass[twocolumn]{article}

\usepackage{graphicx}

\usepackage{url}

\usepackage{endnotes}

\usepackage[symbol]{footmisc}

\title{Paths Not Taken: a Secure Computing Tutorial}

\author{William Earl Boebert\\Retired\endnote{Previously: Honeywell, Secure Computing Corporation, Sandia National Laboratories. \hyphenpenalty=10000 Author contact: boebert@swcp.com\exhyphenpenalty=10000}}
\date{}

\begin{document}

\maketitle

\begin{abstract}

This paper is a tutorial on the proven but currently under-appreciated security mechanisms associated with  "tagged" or "descriptor" architectures. The tutorial shows how the principles behind such architectures can be applied to mitigate or eliminate vulnerabilities.

The tutorial incorporates systems engineering practices by presenting the mechanisms in an informal model of an integrated artifact in its operational environment. The artifact is a special-purpose hardware/software system called a \textit{Guard} which robustly hosts defensive software.  

It is hoped that this tutorial may encourage teachers to include significant past work in their curricula and students who are self-teaching to add that work to their exploration of secure computing.	

\end{abstract}

\section*{Organization of the Tutorial}

This tutorial is divided by topic area to facilitate incorporation in curricula and to assist those who are using it to self-teach. The topic areas are arranged in order of detail, from the most general to the most specific. The first area, Concept of Operations, discusses the issues of systems engineering and the importance of understanding and exploiting the environment in which systems operate. The second section, Models, discusses mental models and the use of abstraction in understanding computer systems. An overview of the Guard model is then given. The Mechanisms topic area is the most specific, where the known technology is described both in detail and in the context of the systems design given in the earlier sections. The final section offers suggestionss to those who may be interested in carrying this work further into an operational system.

\section*{Concept of Operations}
No system can solve every aspect of a given problem, and experience has shown that explicit prior consideration of focus, limits, and approaches will often save time, money, and blood. A major goal of the tutorial is to show how an understanding of the context in which a system operates influences the design of mechanisms. To do so requires presenting a hypothetical operating environment for the hypothetical Guards.

A proven systems engineering \cite{syseng} technique is to define an operating environment and a system's place in it by producing a Concept of Operations, or Conops \cite{conops}. Three Conops-level topics are included to help the student understand the rationale behind the mechanisms: a statement of problem the Guard is intended to solve, the general approach taken for the solution, and the goals of the assurance process. 

\subsection*{Statement of the Problem}

A generic network of interest to attackers is one whose purpose is to provide economic or social value. It does this by running applications on general-purpose platforms which directly or indirectly access the internet. The platforms are typically feature-rich operating systems such as Windows, Linux, or Mac OS. The applications may be a mix of locally developed, purchased, and open source software. Some networks, such as those which control processes such as water supplies and electrical distribution, will be comprised of specialized platforms and applications.

With rare exception, such networks are designed and administered with the applications as a primary concern and defense against attack as secondary. This setting of priorities encourages attackers who already enjoy inherent advantages: attackers can choose the time, place, and nature of an attack and they only need to find one exploitable fault, while defenders must cope with all known and lurking ones, and attacker's motivation is enhanced by the often overlooked fact that for many individuals attack is an adventure while defense is just a job. 

In today's world attackers may also be directly or indirectly protected from retaliation by sovereign states. They will, in general, be adequately funded either by sponsors or by the unregulated transnational money flows made possible by cryptocurrency \cite{tether}. Being funded enables them to be persistent, to perform repetitive attacks and learn from each one. And attackers can exploit features of the internet which permit anonymous action at a distance.\cite{attrib}.  

\subsection*{General Approach}

The sole purpose of a Guard is to host Security Services code. Development of that code is hosted elsewhere. There is also no requirement to support a browser, or data bases, or any other form of productive application. 

The isolation and concentration of Security Services in the Guard nodes means that Guards will be the primary targets of competent attacks, for their defeat  would leave the network largely undefended. The design, implementation, and administration of Guard nodes recognizes this fact. Design and implementation places resistance to attack as a requirement above all others, and administration recognizes that social engineering and supply chain attacks will accompany direct attacks on the network.

Separation of Security Services from applications is motivated by the radically different characteristics of the two classes when viewed from a life-cycle perspective. Application platforms are large, complex bodies of software which are subject to structural decay \cite{belady} unless updated carefully, which means at wide intervals. When new attacks appear, however, Services code must be updated with all deliberate speed so that attackers have the smallest practical window of exploitation. Putting Services code on the same platform leads to one of two undesirable options: if the Services+applications platform is updated at the rate required by Services, the application is at significant risk of decay. If the Services+application platform is updated at the careful pace necessary to prevent decay, updated Services code will be delayed and its effectiveness diluted.

\begin{figure}[h!]
\includegraphics[width=190pt]{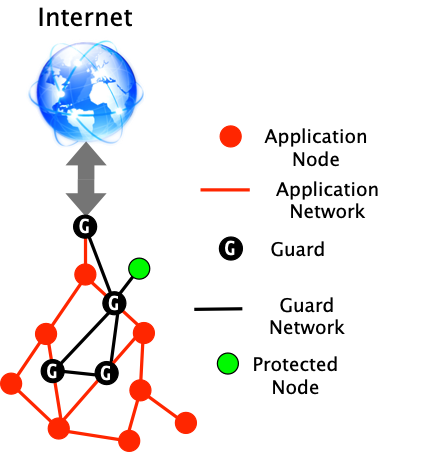}
\centering
\caption{A Network With Guards}
\label{fig:Firewall2}
\end{figure}

Figure \ref{fig:Firewall2} depicts potential applications of Guard instances to network defense.

The most basic application is a traditional firewall that filters packets and raises alarms at the juncture with the internet. Similar tasks could be carried out within the application network to detect or prevent adverse interactions between application nodes. Guards could interact with each other by using out-of-band communication or virtual private network (VPN) technology. A Guard could protect a critical but vulnerable internal node such as a machine learning system for intrusion detection, a network attached storage for backups, or a ``honeypot'' to trap attackers. In all cases the relevant Security Services are determined by the combination of applications and known attacks and therefore system-specific.

The isolation of the Guards from vulnerable network elements is an adaptation of a design principle called \textit{Zero Trust}:

\begin{center}

\textit{The Zero Trust Model is simple: cybersecurity professionals must stop trusting packets as if they were
people. Instead, they must eliminate the idea of a trusted network (usually the internal network) and an
untrusted network (external networks). In Zero Trust, all network traffic is untrusted. }\cite{zerotrust}
	
\end{center}

\subsection*{Assurance}

Assurance determines how human administrators view the cyber systems they administer. Assurance is essential to effective response. Administrators who are assured of known system behavior are more likely to respond decisively and correctly to alarms and other indicators of potential attacks. 

There are two assurance goals associated with a Guard. The first is that the internal mechanisms of the Guard operate as advertised even when attacked. The second goal is to export assurance to the Services code it hosts and from that to the administrators. The second, external assurance goal can be somewhat flippantly expressed as ``WYSIWYG\endnote{``What You See Is What You Get,'' originally a lighthearted way to describe word processors that displayed formatted instead of raw text as you typed.}.'' In more dignified terms, it is the assurance that the semantics of the source code for Services will be enforced by every single instruction executed by the Guard. That assurance enables individuals who are reading the source code for the Services to do so with  confidence. 

These assurance goals determine the nature of the mechanisms in the model and its structuring. More primitive elements are  subject to the greatest amount of testing and analysis. These results then form a basis for arguing that the higher level elements are correct, in the way that lemmas contribute to theorems in mathematical reasoning. 

It should be noted that the assurance exported by the Guard to the Services does not guarantee or even imply that a particular Service is effective. That final, third form of assurance must be provided by those who write and administer the Service. 

It should also be noted that the Guard itself is security policy agnostic. Any rules or restrictions to be enforced on application data are the responsibility of the application-specific security services hosted on the Guard.

\section*{Models}

\subsection*{Models of Nature}

The role of modeling in science was described by Arturo Rosenblueth and Norbert Wiener in the 1940s:

\begin{center}
\textit{No substantial part of the universe is so simple that it can be grasped and
controlled without abstraction. Abstraction consists in replacing the part
of the universe under consideration by a model of similar but simpler structure.}\cite{weiner45}

\end{center}

Similar observations about models were made and expanded by Alan Turing in the preface to his last scholarly paper :
\begin{center}

\textit{This model will be a simplification and an idealization, and consequently a falsification. It is to be hoped that the features retained for discussion are those of greatest importance in the present state of knowledge.}\cite{turing}
\end{center}

Note the depth of insight revealed by Turing's phrase ``it is to be hoped.'' All models are incomplete, and the nature and degree of what is left out is arbitrary. 

\subsection*{Models of Computer Systems}

Turing, Wiener, and Rosenblueth wrote these words before the advent of programmable computers, programmers, and software. That technology requires its practitioners to operate in an arena largely free from physical constraints, as described in a classic passage by Fredrick Brooks:
\begin{center}

\textit{The programmer, like the poet, works only slightly removed from pure thought-stuff. [...] Few media of
creation are so flexible, so easy to polish and rework, so readily
capable of realizing grand conceptual structures.}\cite{Brooks}
	
\end{center}

Brooks goes on to discuss how these characteristics of software make it so difficult to get right. Over the years, practitioners coped with that difficulty by adopting, mostly without explicit thought, an approach that reversed the sequence of scientific modeling: instead of describing an existing tangible object in intangible terms, they create abstract models \textit{in advance} as a way of discussing a largely intangible software/hardware artifact that does not yet exist, and reason about what they are doing in terms of those models. A particular system may be modeled at progressively greater degrees of detail until the point at which a string of ones and zeros is loaded into silicon and activity is generated. In informal terms it is "models all the way down," and the choice given to practitioners is not whether to use a model but rather what kind of model to use.

The model presented in this paper is, as described above, necessarily incomplete. The ``features retained for discussion'' are those which give a Guard the ability to resist attack. These are the mechanisms of \textit{structured memory, demand linking,} and \textit{layer enforcement,} which will be described later. Any hypothetical implementation of a Guard would involve filling in the gaps and manifesting the result in software and hardware. It is the intent of the model that such an activity should require no more than understanding the above elements of known technology.

\section*{The Guard Model}

\begin{figure}[h!]
\includegraphics[width=100pt]{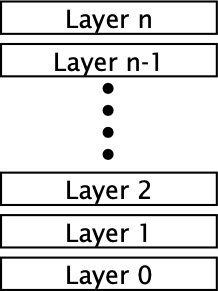}
\centering

\caption{A Layered Structure}
\label{fig:GenLayers}
\end{figure}

The Guard has the general form of a resource-management operating system, and is described in terms of layers of functionality \cite{THE}, as shown in Figure \ref{fig:GenLayers}.

A ``Layer'' in the Guard model is a cluster of data and code objects devoted to a common purpose, such as a file system. When a program in execution (process) has its execution point in procedure belonging to a Layer it is said to be ``in'' the Layer. An object ``belongs'' to a Layer if being there permits execute access (for code objects) or read+write access (for data objects.) These concepts will be discussed in detail later.

Layers are used to apply structure to the complex interactions that exist in resource-management operating systems. The structure, which is motivated by a principle called ``separation of concerns,'' \cite{separation} is intended to improve understanding and support the structured assurance goal described earlier. 

The Guard model carries layering further and provides mechanisms to enforce the Layer structure. These will also be described later.

There a variety of ways in which Layers could interact, such as procedure calls, cooperating sequential processes, or shared data objects. All of these forms of interaction between Layers can be assigned the directional attribute of \textit{dependency.}

Despite it being around for 50 years, there is no industry standard terminology for this relationship. Words like ``depends upon,'' and ``uses'' capture the essence.\cite{parnas} Analyses of dependency are employed in Failure Mode and Effects Analysis (FMEA)\cite{fmea}  and Fault Tree Analysis (FTA)\cite{fta}. In the latter cases the emphasis is on the important question of what happens if a failure occurs in a particular element. 

The term ``influences'' turns the relation around: B influences A if a difference in B causes a difference in A. This paper will employ whichever term leads to the simplest description.

\subsection*{Mechanisms by Layer}

\begin{figure}[h!]
\includegraphics[width=150pt]{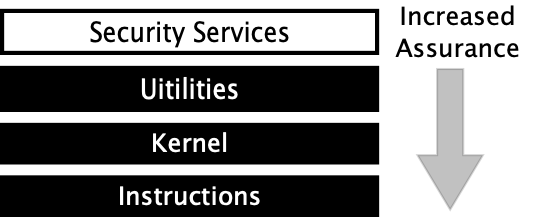}
\centering

\caption{Platform Hierarchy}
\label{fig:layers}
\end{figure}

The platform software can then be separated out and refined into three Layers\footnote{If a common term is capitalized in this paper (e.g. Layer) it denotes the Guard definition; lower case denotes the generic meaning.} as shown in Figure \ref{fig:layers}.

The model assumes that in any hypothetical implementation the amount of testing and analysis (and the resulting assurance) would increase as one goes down the Layers.

The general assignment of mechanism to Layers is as follows:

\begin{list}{}{}

\item\textit{Security Services:} This is the software that provides the network defense. If one characterizes a Guard as an operating system, this software is the equivalent to the applications that run on it. It is assumed that this is a highly dynamic Layer, with modifications made in response to the unannounced arrival of new attacks. 

\item \textit{Utilities:} Mechanisms in this Layer are invoked by procedure calls in the security service code. This Layer includes a file system that associates symbolic names with units of structured memory. It also includes the network interface and things like  event data recording or logging  to preserve and protect event sequences for forensic use. These are well-known functions and accordingly are not treated in detail in the model.

\item \textit{Kernel:} This Layer manages the internal data upon which the proper operation of the Instruction Layer depends. Functions in this Layer are invoked by procedure calls in the  Utilities Layer.

\item \textit{Instructions:} The mechanism in this Layer are invoked by instructions which mimic the interface to conventional hardware with a command, working, and addressing registers. One portion of this Layer provides basic instructions such as arithmetic, test and branch and so forth. The other portion provides security instructions which deal with memory safety and constraints on instruction sequences. \end{list}

\subsection*{Dependencies}

\begin{figure}[h!]
\includegraphics[width=170pt]{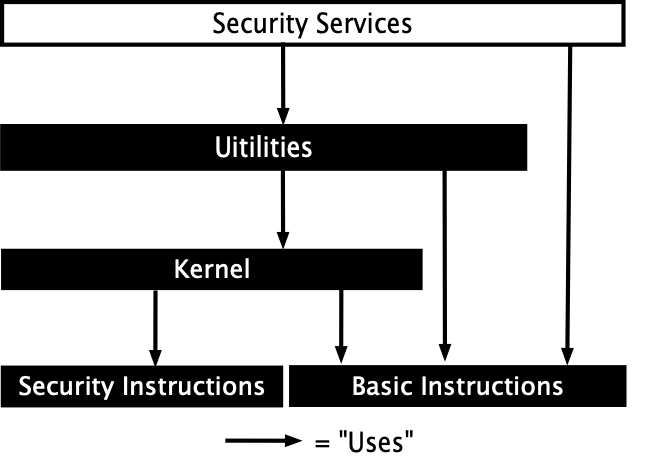}
\centering

\caption{Dependency Restrictions}
\label{fig:depends}
\end{figure}

The dependencies between Layers are shown in Figure \ref{fig:depends}.

This figure shows that the Security Services may use the Utilities and the basic instructions provided by the silicon. The Utilities may use the facilities of the Kernel and the basic instructions. The Kernel, and only the Kernel, is permitted to use the restricted security instructions. These restrictions on dependency are enforced by the mechanisms incorporated in the model.

The dependency restrictions enforce a design principle which states that no element of the Guard shall depend upon an element of lower assurance. This principle supports the structured assurance 
effort.

\subsection*{Operational Context} 

\begin{figure}[h!]
\includegraphics[width=190pt]{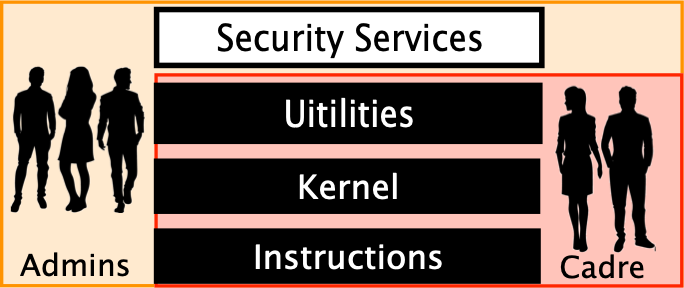}
\centering

\caption{Human and Technical Trust Boundaries}
\label{fig:trusts}
\end{figure}

A basic tenet of systems engineering is that the context in which a system operates must be a factor in its design, and documented in the Conops. In the case of the Guard, that context includes the personnel who build it and those who administer it. These are shown in Figure \ref{fig:trusts}. 

The code that provides the Security Services and the personnel who develop and administer that code form the outer enclave. The code that supports those Services and provides the resistance to attack and the personnel who develop and maintain that code form the inner one. Each enclave grants limited trust to the other.

Service code is reactive in nature and constantly changing, with a premium placed on speed of implementation and updating. To be effective it must be ``quick and dirty'' and therefore assumed to contain errors and possibly subversions. Lower Layer code is carefully planned and stable with a premium placed on assurance, with secrecy of design as a secondary protection. On the other hand, Service code may contain information that the organization using the Guard may wish to hold closely. 

 Outside the administrative enclave all network elements are presumed compromised and hostile, so there is assumed to be a communication channel between administrators and the Guard.  Such a channel would be out-of-band, secured by cryptography and employ some convenient device physically controlled by administrators. This channel is not included in the model, but a general outline of functions and mechanisms may be found in \cite{trustedpath}.


\section*{Mechanisms}

The primary mechanisms of the Guard model are \textit{Structured Memory, Demand Linking}, and \textit{Layer Enforcement}. These were chosen to show how mechanisms can complement each other in a specific operational environment to provide the emergent property \cite{emerge} that the "WYSIWYG" assurance goal of the system is met. Each mechanism is a variant of something that has appeared in one or more predecessor system\endnote{Sources include Multics \cite{multics}, Scomp \cite{scomp}, PSOS \cite{psos}, SAT \cite{sat}, LOCK \cite{smith}, and Sidewinder \cite{sidewinder} The contributions of too many individuals to list are hereby acknowledged; this model is a synthesis and no claim of invention is made or should be implied.}, and has been modeled as simply as possible to illustrate the mechanism's principles. 

The mechanisms are defined in terms of two system primitives: \textit{Segments} and \textit{Processes}. 

\subsection*{Segments}

\begin{figure}[h!]
\includegraphics[width=120pt]{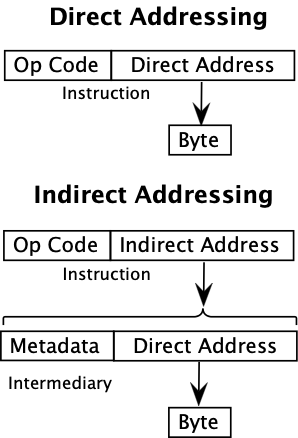}
\centering

\caption{Indirect Addressing}
\label{fig:indirect}
\end{figure}

A Segment is an addressable sequence of bytes. All addressing in a Guard is indirect. 
As shown in Figure \ref{fig:indirect}, places an intermediary element between an instruction's reference to a specific part of memory and the value stored there. 

There are several reasons for using indirect addressing, such as flexibility in managing memory, but the one of greatest significance to resistance to attack is the placing of metadata in the intermediary element. "Metadata," in this context, means data \textit{about} data, security-relevant descriptions such as length or whether the data can be interpreted as instructions. Putting the metadata in the address path provides high assurance that it will uniformly be encountered by the Instruction Layer.

The intermediary, metadata-holding element in the Guard model are \textit{Segment Descriptors}. Individual bytes are addressed by Descriptor, \textit{Offset} pairs.

\subsection*{Processes}

A Process is modeled as an execution point moving through successive object code Segments. Movement from code Segment to code Segment is achieved through the traditional call/return mechanism and pushdown stack.

 Processes begin life in the Services Layer. As a Process moves from code Segment to code Segment it may move from Layer to Layer. A Layer Register is used to track the current Layer the Process is in. The Guard model describes a separate pushdown stack for each Layer.

The Guard model explicitly and deliberately does not include Process muliplexing.

\begin{figure}[h!]
\includegraphics[width=200pt]{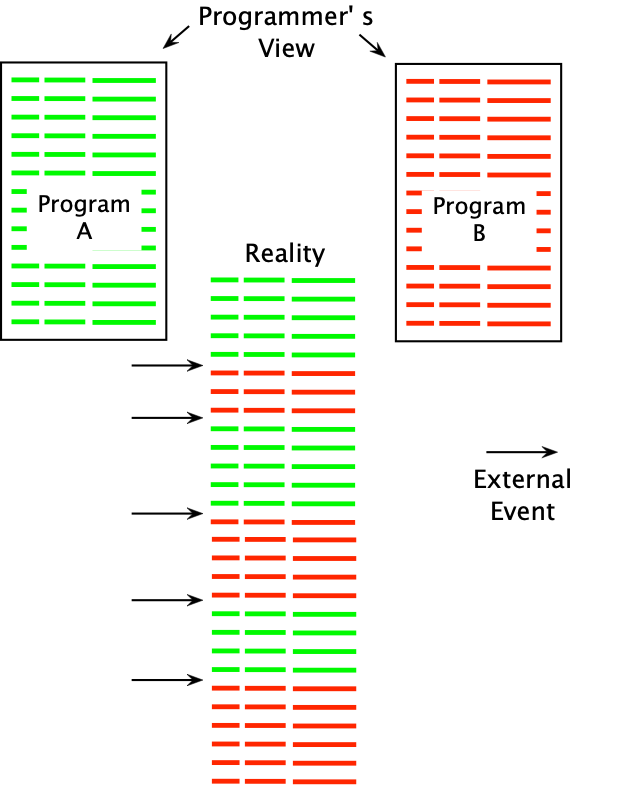}
\centering

\caption{Process Multiplexing}
\label{fig:mux}
\end{figure}

Process multiplexing is the technique dividing multiple program sequences into fragments and interleaving the fragments into a single sequence, as shown in Figure \ref{fig:mux}. 

Simulated parallelism contradicts the ``WYSIWYG'' assurance goal. A programmer or analyst reading a program text will naturally assume that the sequence shown in the text is that which occurs at the Instruction Layer of the Guard. Process multiplexing invalidates that assumption\endnote{The situation is worse if the interleaving is determined by indeterminate external events. Such mechanisms can exhibit the phenomenon called ``Heisenbugs,'' \cite{heisen} where errors manifest themselves in operational use but vanish when instrumentation code is inserted to look for them.}. 

If parallelism is needed then the design approach is to add entire Guards with a shared memory and synchronize them with messages\cite{THE}. This gives each element in the parallel structure the full protection benefit of Layers and ``WYSIWYG'' assurance.

Processes do not execute on behalf of human users and there are no user-settable access rights in a Guard. Guards are not intended to run productive applications like browsers and data bases. As a consequence Guard Processes would, in general, be simple polling loops or linear code sequences that would sleep until woken by an external event. In traditional operating system terms, all Processes are ``daemons'' \cite{daemon}. 

Individual instructions in a Process's sequence have the ability to invoke a special facility called a \textit{Trap}. The Trap mechanism\endnote{Also known as "unprogrammed transfers" or "internal interrupts."} is implemented at the Instruction Layer. It enables a special class of events to be handled at a lower Layer without explicit call/return sequences.

\begin{figure}[h!]
\includegraphics[width=200pt]{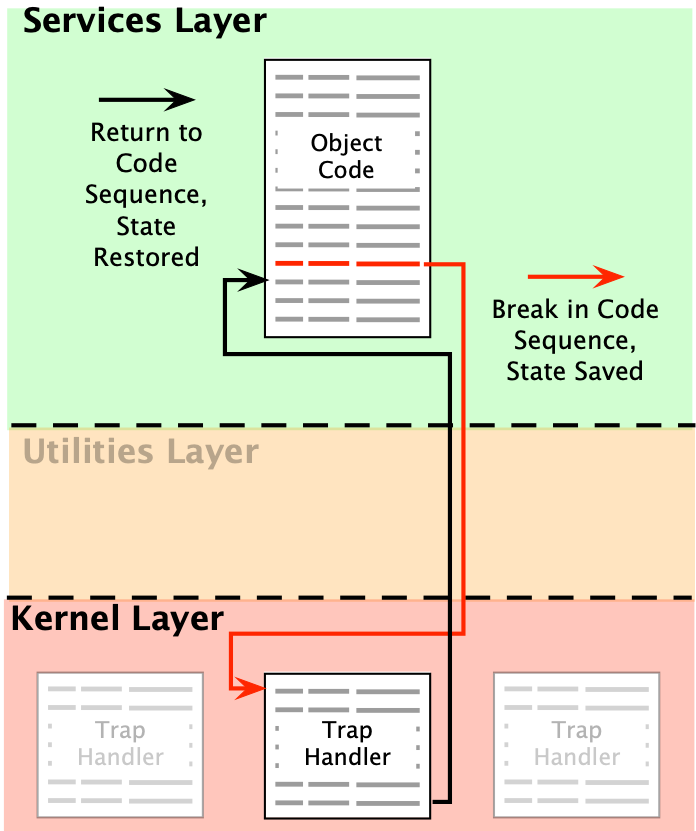}
\centering

\caption{The Trap Mechanism}
\label{fig:trap}
\end{figure}

As shown in Figure \ref{fig:trap}, an Instruction initiating a Trap causes a break in the code sequence, saving of the Process state (register contents and stack) and direct transfer of control to a code segment called a \textit{Trap Handler.} The illustration shows a Trap to the Kernel Layer; Traps to the Utility Layer are also possible.

The kind of Trap initiated determines which Trap Handler is invoked. The Trap Handler performs Trap-specific actions and then restores the Process state and resumes execution at the instruction immediately after the one that initiated the Trap. Trap Handlers thereby become, in effect, software-implemented extensions to instructions\endnote{Because their operation is hidden from programmers or analysts reading code, Traps and Trap Handlers should be used sparingly and with care.}.

Traps can be initiated explicitly by a "Trap" instruction, as a response to an illegal or impossible action such as divide by zero, or when the attempt to fetch a byte encounters a special value.

\subsection*{Structured Memory}

Structured memory is at the heart of the interaction between processes and segments. It has two aspects: the format and meaning of Descriptors, and the manner in which a Descriptor address is mapped onto the conventional primary and secondary memories at the Instruction Layer\endnote{The assumption here is that the Instruction Layer is eventually embodied in hardware}.

\subsubsection*{Descriptors}

\begin{figure}[h!]
\includegraphics[width=170pt]{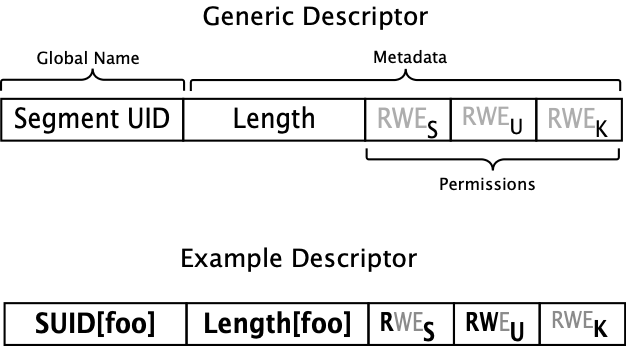}
\centering

\caption{Descriptors}
\label{fig:Descriptors}
\end{figure}

The generic form and an example of a Descriptor is shown in Figure \ref{fig:Descriptors}. Descriptors have two main fields, a \textit{Segment UID} or \textit{SUID} which uniquely identifies the Segment, and a set of metadata. The inclusion of metadata makes the Guard model an example of a ``tagged architecture,'' \cite{tagged} where the tags apply to segments rather than words or bytes. 

The metadata consists of a length field, whose inclusion and checking by the mapping logic makes memory safety \cite{memsafe}	automatic, and a set of permissions. There is one permissions field for each Layer. The values shown (read, write execute) are examples and other values, such as ``append'' are possible. There is no manual setting of permissions in the model; all permissions are determined at the time the Guard software is initialized prior to installation. 

 When permission violations are detected by the mapping logic, it generates a Trap to an error Trap Handler which will perform appropriate actions like sending an alarm to administrators and halting.
 
The example Descriptor in Figure \ref{fig:Descriptors} is for a data Segment called ``foo,'' which will be used as an example throughout the discussion of mechanisms. It is assumed to be managed by a Utilities Layer routine called ``foo\_owner.''

 The ``SUID[foo]'' value denotes its identity and ``Length[foo]'' its size. It is a Utility Layer Segment, and this is shown in its permissions. The first permission field says that a Process executing in the Services Layer has ``read'' only access to ``foo,'' one in the Utilities Layer has ``read'' and ``write,'' and one in the Kernel Layer is denied all access. This would be a typical set of permissions for something like a file system Segment.

\subsubsection*{Descriptor Addressing}
\begin{figure}[h!]
\includegraphics[width=210pt]{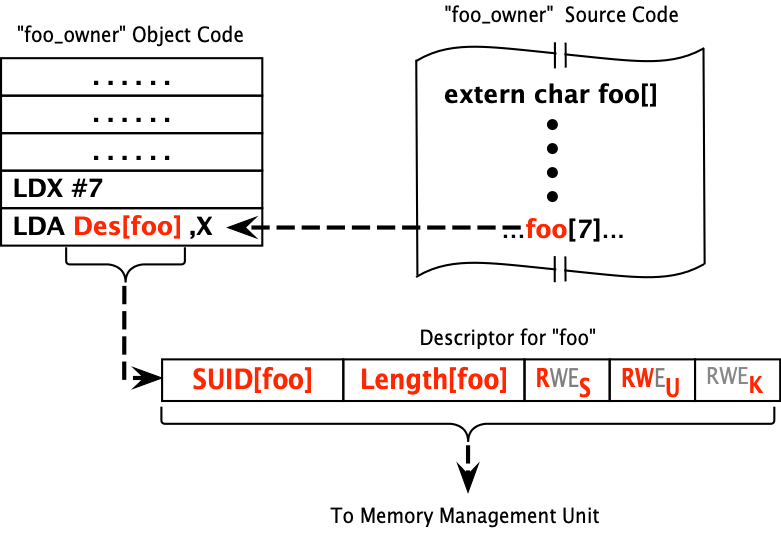}
\centering

\caption{Ideal Name Association}
\label{fig:Desired}
\end{figure}

Figure \ref{fig:Desired} continues the example by showing the desired relationship between the three levels of names that connect source code to  physical memory. The source program for ``foo\_owner'' contains the declaration of ``foo'' and a reference to a byte at the 8th position in ``foo'' by means of an indexed instruction:
\begin{list}{}{}
	\item LDX \#7: Load index register with constant ``7''
	\item LDA Des[foo],X: Load accumulator with byte at physical address defined by the Descriptor for foo offset by contents of index register.
\end{list}

The reference is then sent to an entity called the \textit{Memory Management Unit} which will perform a mediated mapping to physical memory in a manner described later.

\begin{figure}[h!]
\includegraphics[width=210pt]{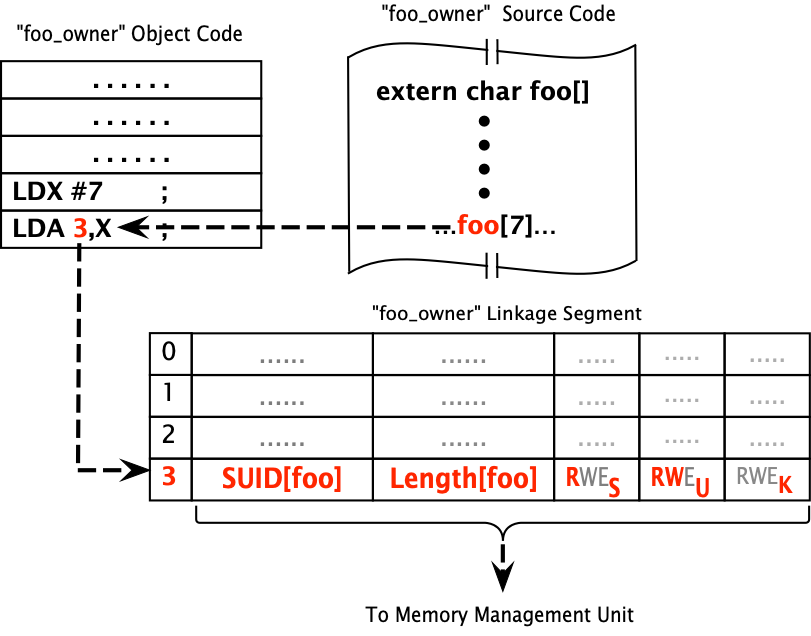}
\centering
\caption{Descriptor in Linkage Segment}
\label{fig:WithLS}
\end{figure}

The addressing mechanism achieves the above ideal association by adding a Process-specific Segment called (again, for historical reasons) the \textit{Linkage Segment} and having the compiler insert a local address within the Linkage Segment in the LDA instruction. (Figure \ref{fig:WithLS}) Local address 0 will, by convention, point to a ``scratch'' Segment containing local variables.

The manner in which Linkage Segments are constructed will be described below. For now it is sufficient to assume it happens correctly and consider the logic that maps a descriptor to a physical memory address.

\subsubsection*{Memory Management Unit}
The Memory Management Unit, or \textit{MMU}, is implicitly invoked whenever an address appears in one of the Instruction Layer registers. The Memory Management Unit is assumed to move segments between primary and secondary memory using any of a variety of well-known techniques (paging, paged segments, fixed segments) and is not further described here. It likewise is assumed to use known caching strategies to exploit locality of reference.

\begin{figure}[h!]
\includegraphics[width=210pt]{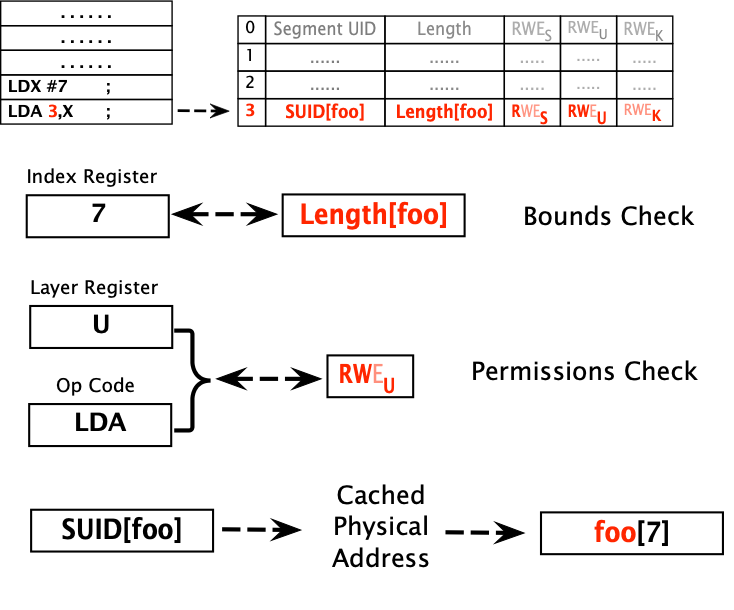}
\centering
\caption{MMU Checks Metadata}
\label{fig:MMU}
\end{figure}

Placing the MMU between the Instruction Layer and the physical memory means that it is uniformly invoked at each memory access. That invocation provides the opportunity to check the attempted instruction against the metadata contained in the Descriptor (Figure \ref{fig:MMU}) before executing it. First, the bounds check is made by comparing the indexing value (Offset) to the length of ``foo.'' Then the Layer Register is used to retrieve the relevant permissions and those are checked against the required permission of the op code; in this case a ``fetch'' instruction so ``read'' access is required. If either of those checks fail, a Trap to an error Trap Handler occurs. Otherwise, the SUID is used to index internal MMU caches of SUID/physical address pairs to locate the segment containing the byte. This physical address is then combined with an Offset to send the byte to the designated working register.

\subsubsection*{Observations} Inherent safety of memory makes the Guard resilient against malfunctioning or malicious code at the Services Layer. That resilience would permit service code to be produced quickly by programmers or generative AI, which would then increase responsiveness to new or ``zero day'' attacks. Mistakes or exploits that would enable takeover of an entire platform based on unstructured memory would simply generate a Trap into an error Trap Handler.

\subsection*{Demand Linking}

When originally formulated for the Multics \cite{multics} system, this mechanism was called ``dynamic linking.'' That term has been pre-empted over time by a different mechanism and so the  more descriptive term ``demand linking'' is used here.

Linking is the process of setting up the relations shown in Figure \ref{fig:WithLS}. In the Guard model it is deferred until the first time Process makes reference to a Segment, hence it happens ``on demand.''

\begin{figure}[h]
\includegraphics[width=0.45\textwidth]{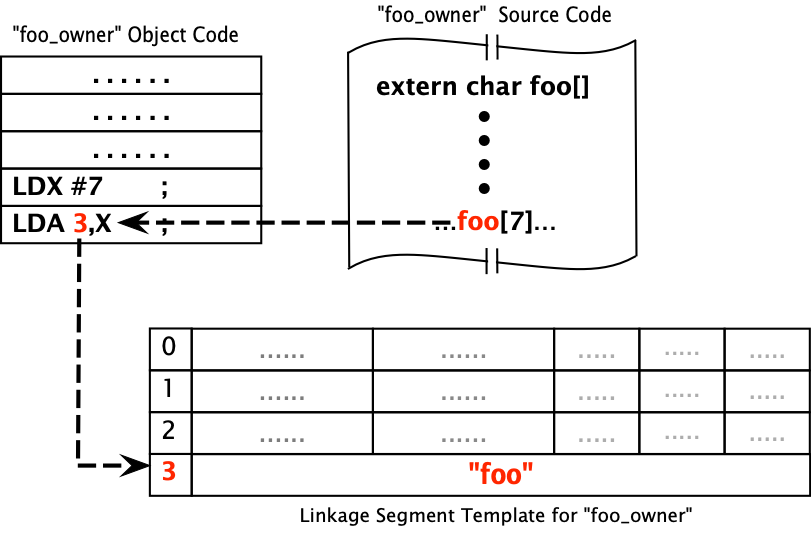}
\centering

\caption{Linkage Segment Template.}
\label{fig:PreLink}
\end{figure}

The linking sequences begins with the translation of source to object code, which as noted above, takes place on a separate development platform. The compiler produces two segments for each unit of source code: an object code segment, which contains local addresses as described previously, and a \textit{Linkage Segment Template}  (Figure \ref{fig:PreLink}) which is initialized with the symbolic names of external segments. The local addresses incorporated in the object code are indexes into the template.

When a Process is initialized a copy of the Linkage Segment is made from the template and associated with the (Process, object Segment) pair. This new Linkage Segment will initially be filled with symbolic names as shown in Figure \ref{fig:PreLink}.

\begin{figure}[h]
\includegraphics[width=0.45\textwidth]{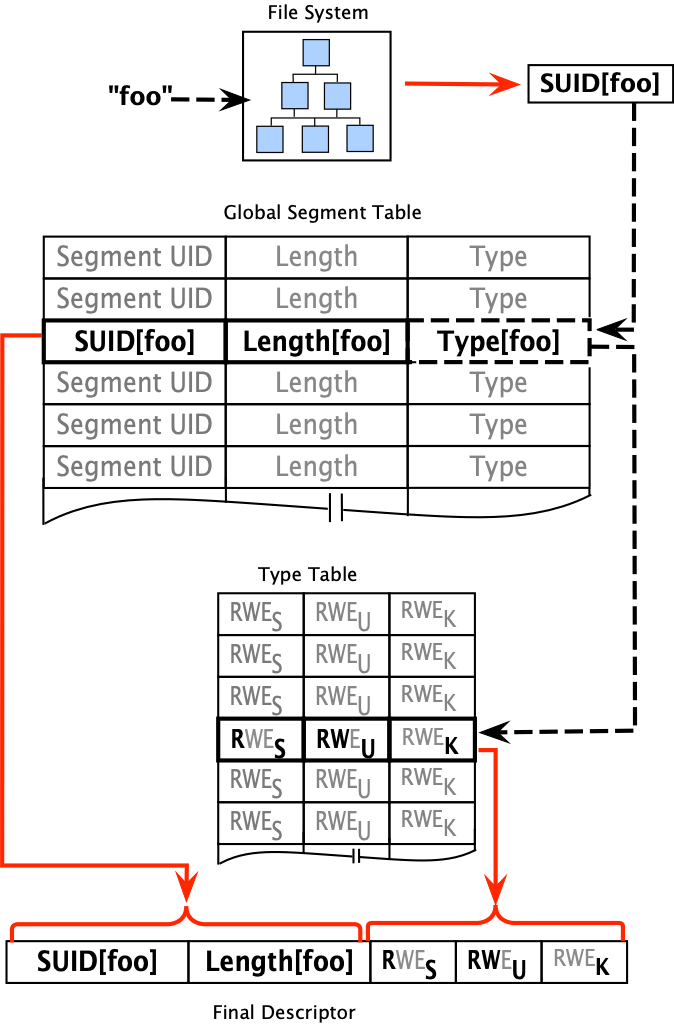}
\centering

\caption{The Linking Process.}
\label{fig:Linking}
\end{figure}

When the LDA 3,X instruction is first encountered by the Instruction Layer, that logic will fetch element 3 of the linkage segment and encounter the symbolic name ``foo'' instead of a Descriptor. This will force a Trap to a Utilities Layer Trap Handler which will start the sequence shown in Figure \ref{fig:Linking}.

The Utilities Layer routine will use the file system to extract the Segment UID associated with the symbolic name ``foo.'' It will then call a Kernel Layer routine which will extract a partial Descriptor for ``foo'' from a 
\textit{Global Segment Table} or \textit{GST}. The Segment UID and Length fields of the GST will be used to construct the first two fields of foo's Descriptor. The Type field of the GST will be used to extract the per-Layer permissions from a \textit{Type Table} and they will complete the Descriptor, which will replace the symbolic name ``foo'' in the Linkage Segment and leave it as shown in Figure \ref{fig:WithLS}. The routines will then ``unwind'' back to the Services Layer, where the LDA 3,X instruction will be repeated and this time trigger an access through the MMU.

\begin{figure*}
\centering	
\includegraphics[width=0.9\textwidth]{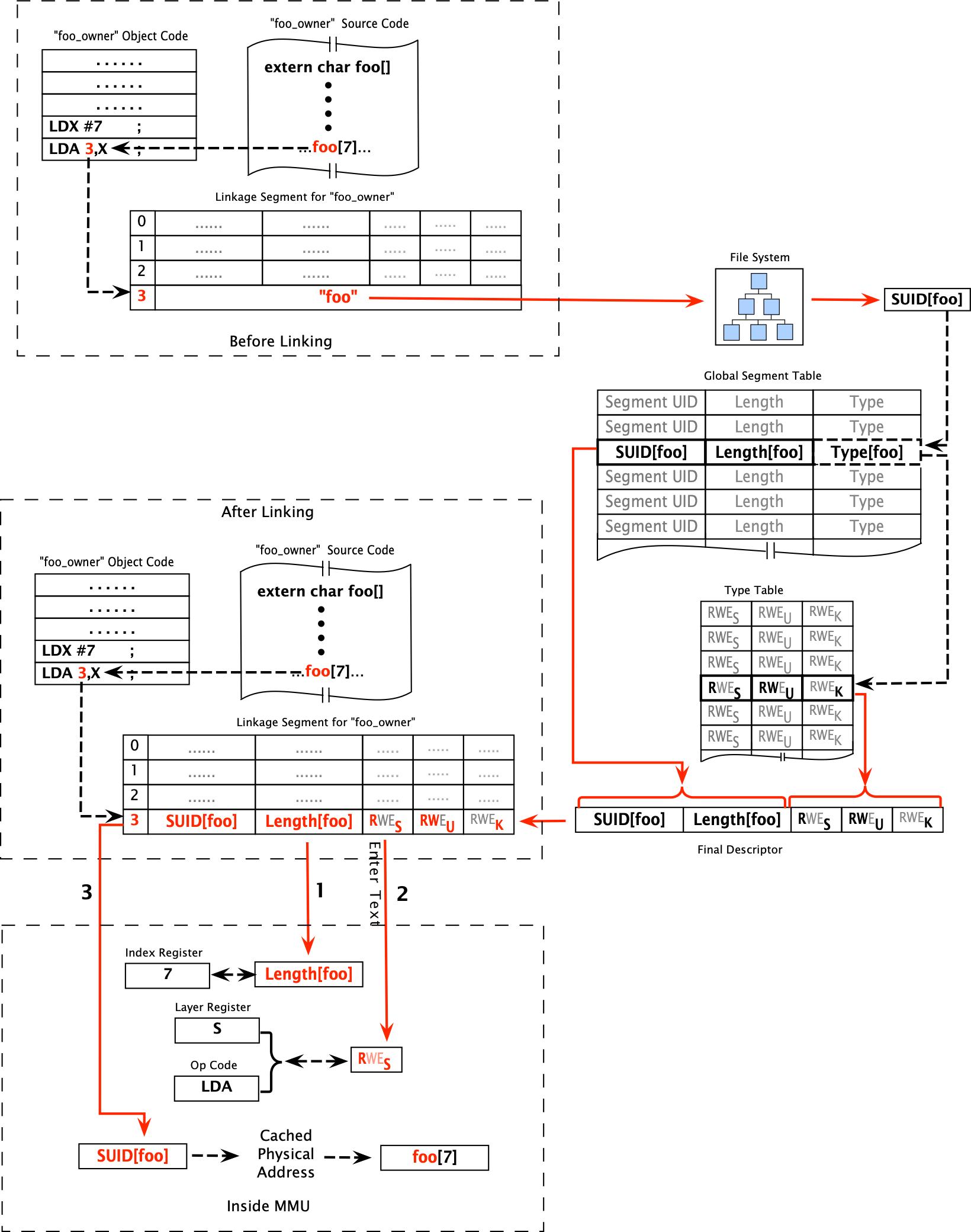}
\caption{From Source Code to Physical Storage}
\label{fig:linked}
\end{figure*}

The full linkage association from symbolic name to the storage subsystem is shown in Figure \ref{fig:linked}.

\subsubsection*{Observations}

The linking mechanism makes use of indirection and Traps instead of explicit procedure calls\endnote{Indirection and Traps carry a performance penalty, which historically has inhibited their acceptance in application system. This inhibition does not apply to a dedicated security system.}. This feature, along with separate Linkage Segments for each Process, enables a Guard to update security services without interruption. Assume that a service, say ``firewall'' has been running for some time.If it is necessary to replace it, administrators can change its name to ``oldfirewall" without affecting its operation: all linkage tests and actions have been performed.  The new version then can be installed with the name ``firewall'' and as it executes its links will be resolved on demand. Eventually the ``oldfirewall" service code will fall into disuse and it can be deleted. Such a facility will enable new and updated Services to be installed as soon as they are ready.

Demand linking also minimizes the number of completed links that are available to malicious or malfunctioning code, a common vulnerability in other linking approaches. Programmers can have a tendency to include whole libraries on a ``just in case'' basis. If these libraries are linked together as a large ``furball'' of code, vulnerability exploits can theoretically go anywhere inside that code set. Demand Linking only links Segments that are actually referenced, and has the ability to produce event data records that can record malicious or malfunctioning code attempts to improperly access Segments.

\subsection*{Layer Enforcement}

The Layer Enforcement mechanism supports the assurance of the Guard itself by enforcing the separation of concerns and dependency restrictions upon which that assurance is based. It is a run-time control mechanism which alters the permissions available to a Process as it moves from Layer to Layer in the course of execution.

Each of the three upper Layers has a dedicated stack. Stacks contain only Descriptor,Offset pairs; no data is pushed onto the stacks. Only Instruction Layer code is permitted to operate on stacks. This restriction insures that all memory references are mediated by the MMU, and eliminates the classic ``stack overflow'' vulnerability.

The distinguishing characteristic of Layer Enforcement is that permissions are both gained and relinquished as a Process moves between Layers. Properly configured, this enforces unidirectional dependency relationships which support structured assurance.

\begin{figure}[h]
\includegraphics[width=0.45\textwidth]{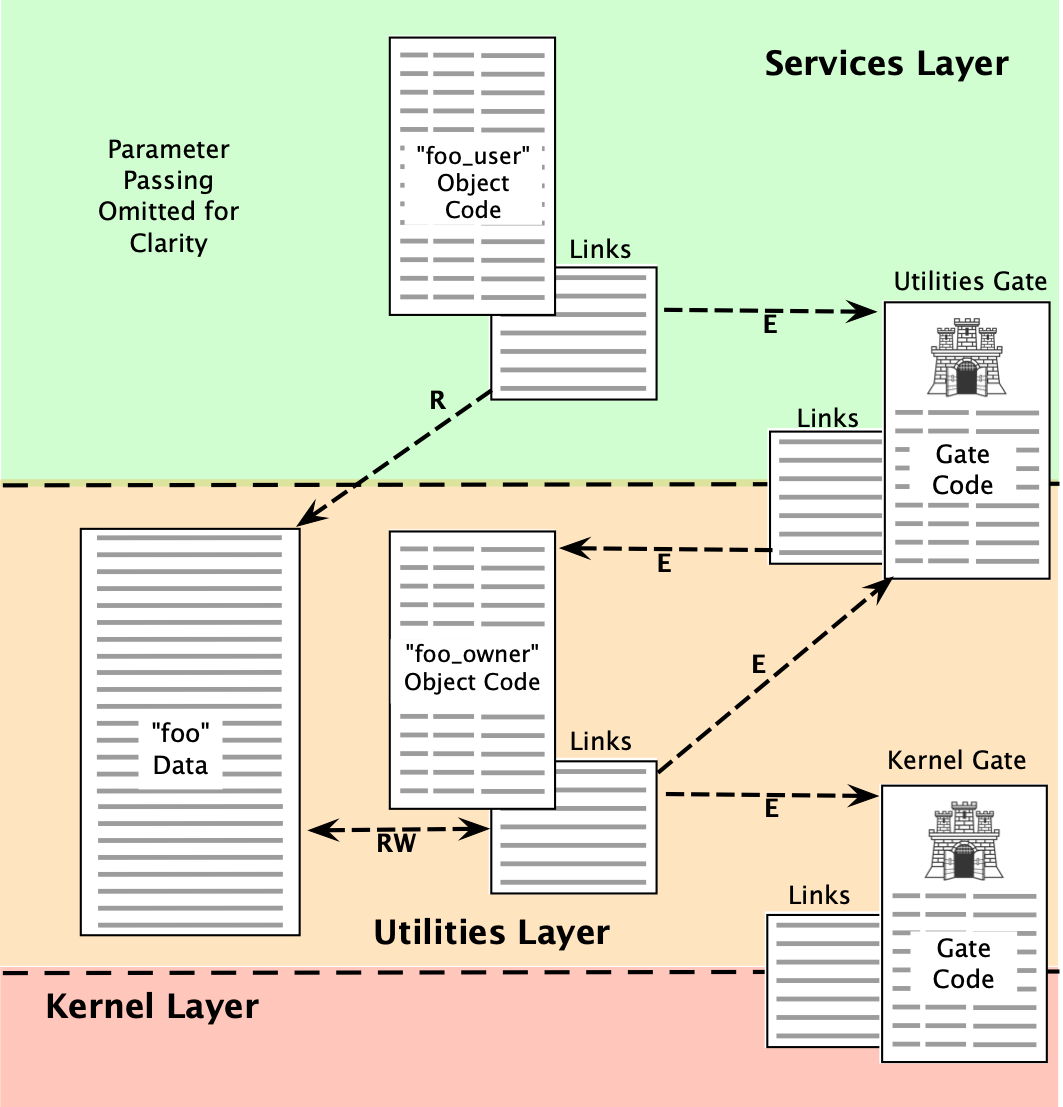}
\centering

\caption{Layers and Gates}
\label{fig:Gate1}
\end{figure}

Figure \ref{fig:Gate1} shows a characteristic arrangement of object code Segments surrounding the data Segment ``foo'' used in the earlier examples. Segment ``foo'' is managed by a Utilities Layer routine called ``foo\_owner,'' whose responsibility it is to validate and implement requests for changes to ``foo.'' Segment ``foo'' is used by a Services Layer routine ``foo\_user,'' which can access but not modify its contents.  Layers are accessible through distinguished object segments called \textit{Gates} which have the restricted ability to change the value of the Layer Register and the stack, as well as insuring proper sequence of transition during execution of call and return instructions. 

 There are two Gates, one which forms the entry to the Utilities Layer from the Services Layer and a second for entering the Kernel Layer from the Utilities Layer. No Kernel Layer function is involved in the example and its Gate and the associated link is included only for completeness. The Instruction Layer, as noted above, is invoked implicitly by changing values of the Instruction Register.

The execution sequence of a Process requesting manipulation of ``foo'' begins with a call from ``foo\_user'' to the Utilities Gate. That Gate in turn changes the value of the Layer Register and calls ``foo\_owner,'' which performs the requested operation and then returns back through the Gate to the Services Layer.

Figure \ref{fig:Gate1} shows the segments and links as if all were visible to all, a circumstance that would never occur in actual operation. 

\begin{figure}[h]
\includegraphics[width=0.45\textwidth]{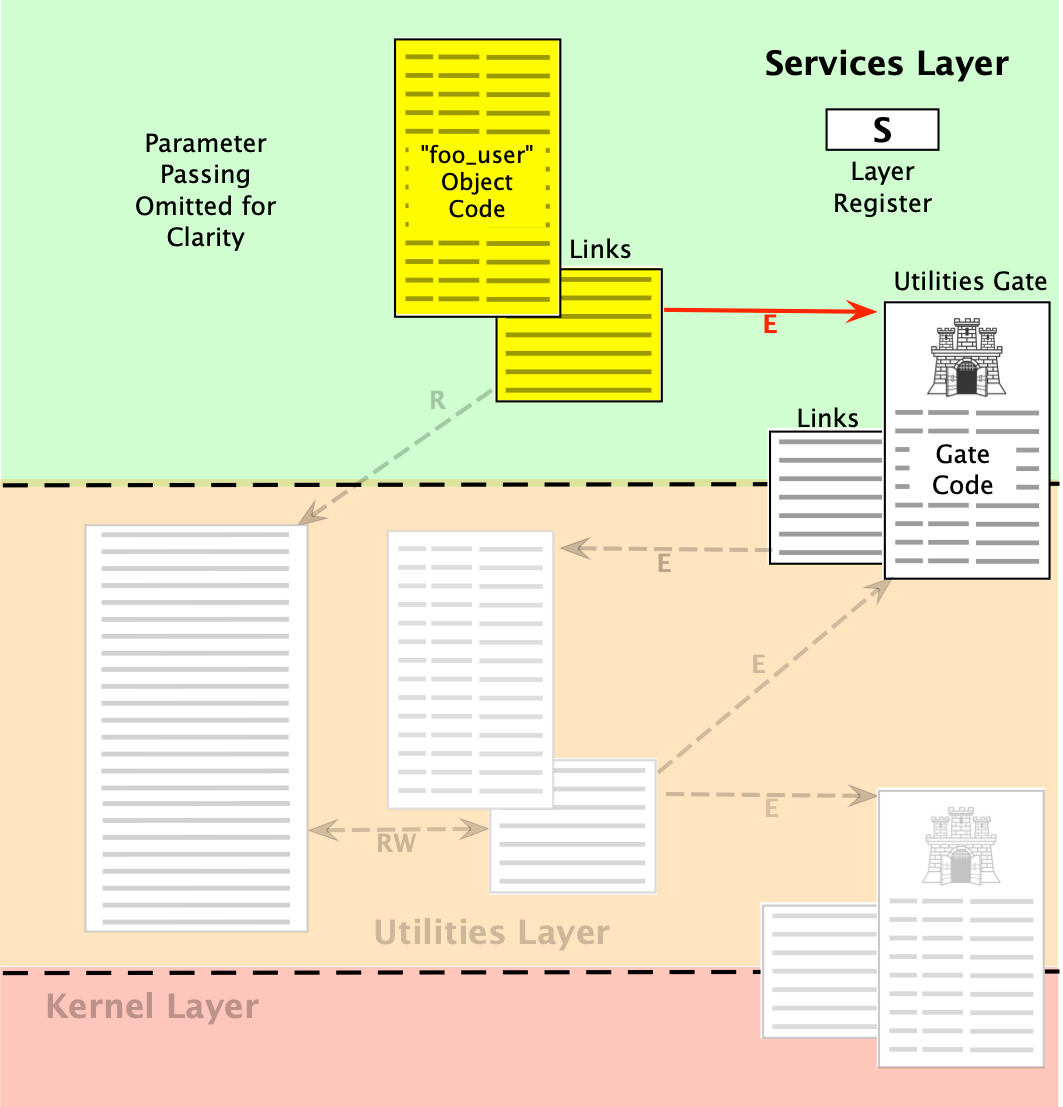}
\centering
\caption{Process in Services Layer}
\label{fig:Gate2}
\end{figure}

The permissions when the execution point of the Process is in the Services Layer object code segment ``foo\_user'' (shown in yellow) provide the visibility shown in Figure \ref{fig:Gate2}. The Segment ``foo\_user'' is exercising its link to the Utilities Gate while the Layer Register is set to ``S'' for Services. 

\begin{figure}[h]
\includegraphics[width=0.45\textwidth]{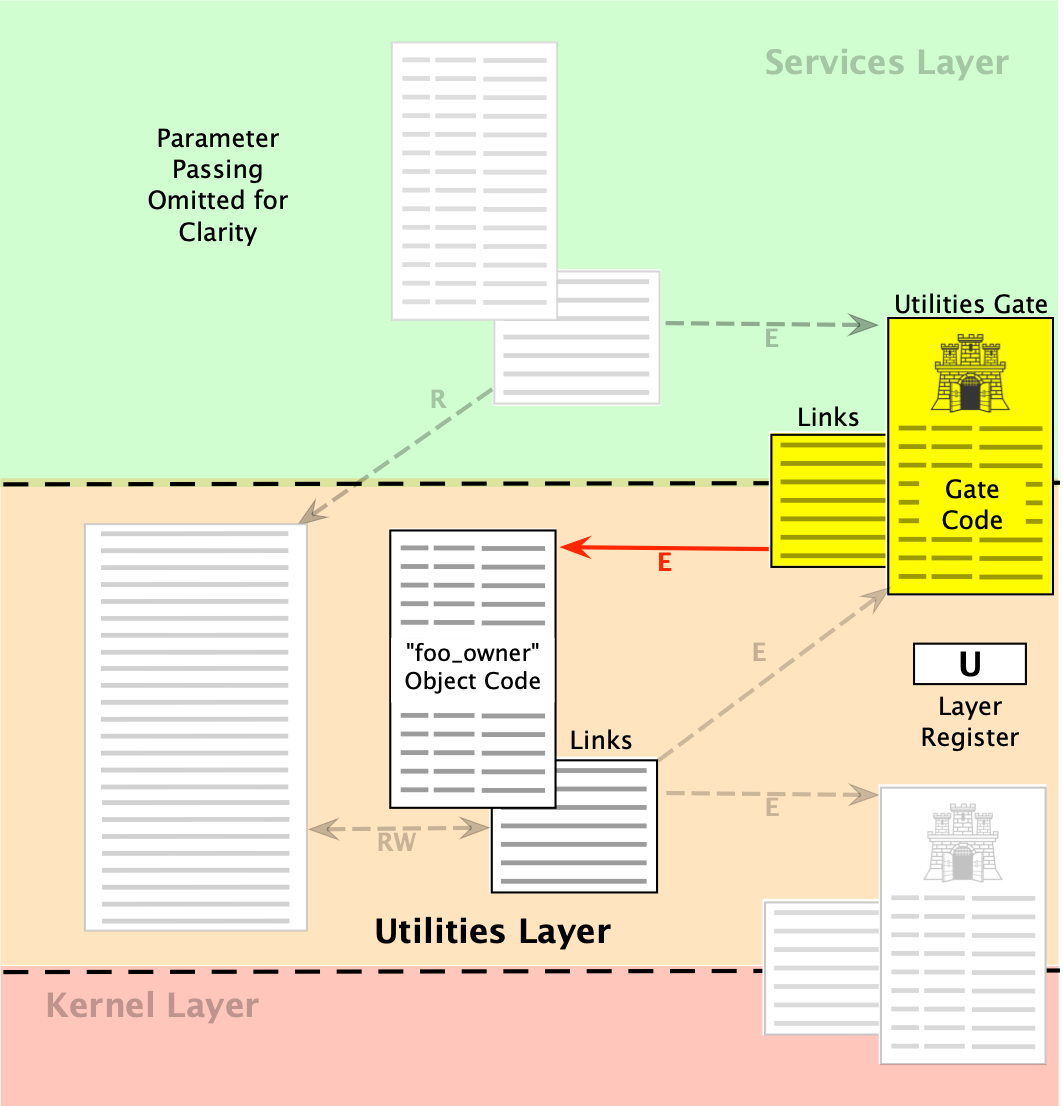}
\centering
\caption{Process Executing Gate}
\label{fig:Gate3}
\end{figure}

When the execution point of the Process is in the Gate shown in Figure \ref{fig:Gate3}, that Gate code will change the value of the Layer Register to ``U'' for Utilities and execute a call to ``foo\_owner.'' At this point the Gate code is prevented from depending upon (e.g., by calling)  ``foo\_user'' or any other Services routine by the configuration of permissions, thereby preventing attacks based on maliciously or erroneously malformed parameter sets from ``foo\_user.''

\begin{figure}[h!]
\includegraphics[width=0.45\textwidth]{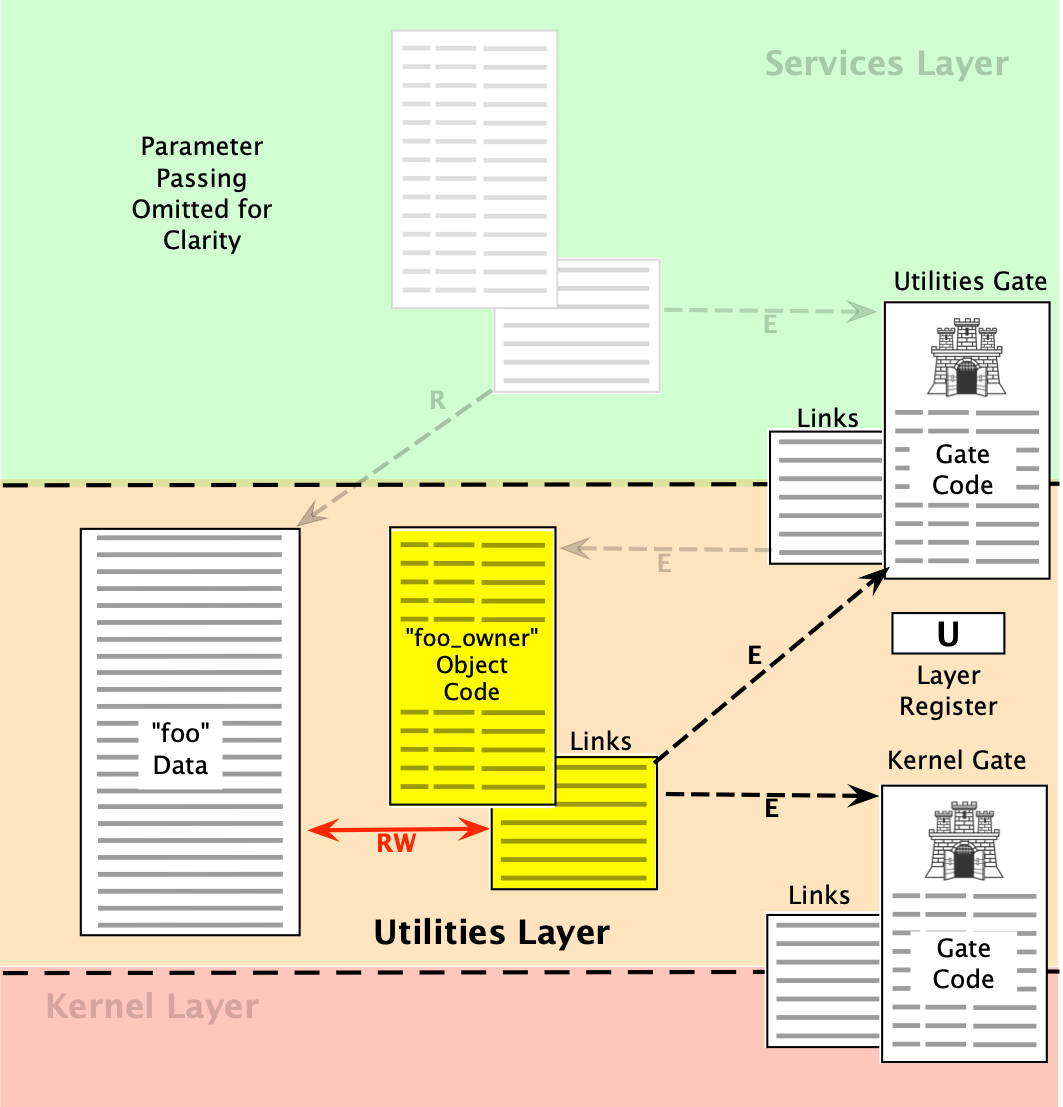}
\centering
\vspace{20pt}
\caption{Process in Utilities Layer}
\label{fig:Gate4}
\end{figure}

Finally, the Process executes the object code Segment ``foo\_owner'' in the Utilities Layer, giving the accesses shown in Figure \ref{fig:Gate4}. At this point ``foo\_owner'' has the option of returning back through the Utilities Gate to the Services Layer or initiating a crossing into the Kernel Layer for some restricted function, such as changing the size of ``foo.''

\subsubsection*{Observations}

The repetitive pattern of permissions shown in the example explains why the permission fields are kept in a separate Type Table rather than entered on a per-descriptor basis in the Global Segment Table. Types are essentially equivalence class of permission, and grouping them as such in the Type Table simplifies an inherently error-prone configuration process and enables modifications to be made by changing a single entry rather than having to examine every descriptor in the Global Segment table to determine if the change was relevant to it.

There is an alternate mechanism in which Gates are initiated by a Trap when a link with execute permissions crosses to a new Layer, as would occur if the Segment ''foo\_user'' were directly linked to ``foo\_owner.'' The mechanism involving explicit calling of the Gate Segment by ``foo\_user'' was chosen in the interests of the simplest explanation of the principle that permissions are both gained and relinquished. 
\newpage

\section*{Notes on Further Work}

It is clearly feasible for students who wish to pursue this alternative approach to do so, even to the point of attaching a running system to the internet and watching it being attacked. Inexpensive x86 machines of sufficient power are available within a student budget. Two such machines, one as a development platform and the other as a target would be enough to support even a team of students. Filling in the omissions in the model presented here with executable code would provide experience in machine-level programming, integration, and assurance in a structured program with defined goals.

A logical plan would be to first construct a virtual machine \cite{vm} beginning with a Memory Management Unit and then an Instruction Layer to provide a suitable register and command set\endnote{If the results of the effort looked promising, elements such as a MMU coprocessor could be made with a  a field programmable gate array \cite{fpga}.}. The next step would be to adapt a suitable compiler to generate Linkage Segments as well as code for the virtual machine. This could be tested with hand-generated Descriptors until Demand Linking was built. The last step in producing a basic platform would be Layer Enforcement. 

Once a platform had been built the students could gain experience in vulnerability assessment by subjecting it to penetration tests and eventually connecting it to the internet to observe real-world attacks.

\section*{Acknowledgements}

The author would like to thank Tom Berson and Sebastian Frazier for their comments on an earlier draft of this paper.


\newpage
\theendnotes

\end{document}